\documentclass[prl,twocolumn,showpacs,superscriptaddress,byrevtex]{revtex4}

\usepackage{graphicx}
\usepackage{dcolumn}
\usepackage{amsmath}
\usepackage{epsfig}
\usepackage{bm}
\usepackage{amssymb}

\newcommand{\ket}[1]{\ensuremath{|\,{#1}\,\rangle}}
\newcommand{\bra}[1]{\ensuremath{\langle\,{#1}\,|}}
\newcommand{\braket}[2]{\ensuremath{\langle\,{#1}\,|\,{#2}\,\rangle}}

\newcommand{\op}[1]{\ensuremath{\hat{#1}}}

\newcommand{\lsub}[1]{\ensuremath{_{_{\!\scriptstyle #1}}}}
\newcommand{\ce}[1]{\ensuremath{\mathcal{#1}}}
\newcommand{\ds}{\displaystyle}
\newcommand{\st}{\scriptstyle}

\newcommand{\itg}[1]{\ensuremath{\int\!\!d{#1}\!\!}}
\newcommand{\itgf}[1]{\ensuremath{\int\!\!d{#1}\,}}

\newcommand{\sinc}{\ensuremath{\mbox{\hspace{1.3pt}sinc}\,}}

\begin{document}


\title{Generation of maximally entangled states of qudits using twin photons}

\date{\today}

\author{Leonardo Neves}
\author{G. Lima}
\affiliation{Departamento de F\'{\i}sica, Universidade
Federal de Minas Gerais. Caixa Postal 702,
Belo~Horizonte,~MG 30123-970, Brazil.}

\author{J. G. Aguirre G\'omez}
\affiliation{Departamento de F\'{\i}sica, Universidade
Federal de Minas Gerais. Caixa Postal 702,
Belo~Horizonte,~MG 30123-970, Brazil.}
\affiliation{Center for Quantum Optics and Quantum Information,
Departamento de Fisica, Universidad de Concepci\'on,  Casilla 160-C, 
Concepci\'on, Chile.}

\author{C. H. Monken}
\affiliation{Departamento de F\'{\i}sica, Universidade
Federal de Minas Gerais. Caixa Postal 702,
Belo~Horizonte,~MG 30123-970, Brazil.}

\author{C. Saavedra}
\affiliation{Center for Quantum Optics and Quantum Information,
Departamento de Fisica, Universidad de Concepci\'on,  Casilla 160-C, 
Concepci\'on, Chile.}

\author{S. P\'adua}
\email{spadua@fisica.ufmg.br}
\affiliation{Departamento de F\'{\i}sica, Universidade
Federal de Minas Gerais. Caixa Postal 702,
Belo~Horizonte,~MG 30123-970, Brazil.}



\begin{abstract}
We report an experiment to generate maximally entangled states of
$\mathcal{D}$-dimensional quantum systems, \textit{qudits},
by using transverse spatial correlations of  two parametric  down-converted
photons. Apertures with $\mathcal{D}$-slits in the arms of the twin photons
define the qudit space. By manipulating the pump beam correctly the twin
photons will pass only by symmetrically opposite slits, generating entangled
states between these different paths. Experimental results for qudits with
$\mathcal{D}= 4$ and 8 are shown. We demonstrate that the generated states
are entangled states.
\end{abstract}

\maketitle



The interest in studying higher dimensional entangled states comes both from
the foundations of quantum mechanics and from the development of
new protocols in quantum communication. For instance, it was
demonstrated that maximally entangled states of two quantum systems in a
$\mathcal{D}$-dimensional Hilbert space, \textit{qudits}, violate local 
realism stronger than qubits \cite{zeilinger1}. 
Entangled qudits are more resistant to noise than qubits, as was shown 
in~\cite{zeilinger1,collins}.
In quantum cryptography \cite{ekert}, the use of entangled qutrits
($\mathcal{D} = 3$) \cite{peres,durt} or qudits \cite{bjork,gisin2} instead 
of qubits is more secure against eavesdropping attacks.
Moreover, one knows that the protocols like quantum 
teleportation~\cite{bennett} or quantum cryptography~\cite{ekert}, work best 
for maximally entangled states.
All these facts motivate the development of techniques to  generate
entangled states among quantum systems in higher dimensional Hilbert space
with a good quality of entanglement.

Recently, spontaneous parametric down-conversion (SPDC) has been
used for realizing  entangled qudits. In Ref.~\cite{howell}
four polarization entangled photons are used to obtain two entangled qutrits.
The use of two photons in higher dimensional space is another possibility.
Entangled qutrits with two photons using an unbalanced 3-arm fiber optic
interferometer \cite{gisin4} or photonic orbital angular momentum
\cite{zeilinger2} has been demonstrated.
Time-bin entangled  qudits up to $\mathcal{D} = 11$ from pump
pulses generated by a mode-locked laser  has also been reported \cite{gisin3}.

In this letter, we demonstrate the experimental generation of
\emph{maximally} entangled states of qudits by using the transverse spatial
correlations of the photon pairs (\emph{biphotons}) produced by SPDC.
Biphotons are sent through apertures with $\mathcal{D}$-slits. The  
$\mathcal{D}$ possible paths (slits) followed by each photon of the pair are 
defined as our qudit space. Due to a transference of information from the 
pump laser beam to the two-photon state \cite{monken1}, we can control the 
transverse correlations of the photon pairs passing by the slits, by 
manipulating the pump beam. A proper manipulation of this one allow us to 
make the biphotons pass only by symmetrically opposite slits, generating 
entangled states between these different transverse spatial modes.
Results for qudits with $\mathcal{D} = 4$ and 8 are shown and
the scheme described here can be extended to higher dimensions.
We give a brief theoretical description of this process and present the
experimental results and discussion.


Here, it is sufficient to write the equations in one dimension.
Considering the degenerate case and using the monochromatic,
paraxial and thin crystal approximations, the two-photon state
generated by SPDC and transmitted through apertures is given by \cite{artleo}
\begin{eqnarray}   \label{Psi}
\ket{\Psi} & \!\!\!\!\! = \!\!\!\!\! & \lambda \! %
\itg{q_{1}}\itg{q_{2}}\itg{x_{1}}\itgf{x_{2}}
e^{i\frac{k}{8z_{A}}(x_{2}-x_{1})^{2}}
e^{-i(q_{1}x_{1} + q_{2}x_{2})} \nonumber \\[2mm]
&  &  \times
A_{1}(x_{1}) A_{2}(x_{2})  W\!\bm{(}{\st\frac{1}{2}}(x_{1}+x_{2});z_{A}\bm{)}
\ket{1q_{1}}\ket{1q_{2}},
\end{eqnarray}
where $q_{j}$ and $x_{j} $ are the transverse components of the wave vector
and position, respectively, of the down-converted photons in modes
$j=1,2$; $A_{j}(x_{j})$ is the transmission function of the aperture in mode
$j$ and $W(\xi ;z_{A})$ is a function that describes the pump beam transverse
field profile at a longitudinal distance $z_{A}$ from the crystal
(plane of the apertures).
We see that the pump beam transfers an information to the two-photon
state. Defining the apertures, we can modify the transverse
spatial correlations of the twin photons by modifying the function
$W(\xi ;z_{A})$, which is achieved by  manipulating the pump beam.

Suppose that each photon component of the down-converted pair is sent to a
$\ce{D}$-slit ($\ce{D}\geq 2$) in $z=z_{A}$. The transmission function
of this aperture is
\begin{equation}  \label{slits}
A_{j}(x_{j}) = \sum_{l=-l_{D}}^{l_{D}} \Pi \! %
\left( \frac{x_{j} - ld}{2a}\right),
\end{equation}
where $j=1,2$, $l_{D}\equiv (\ce{D}-1)/2$ and $\Pi (\xi)$ is a rectangle 
function; $a$ is  the slit half width and $d$ is the separation between two 
consecutive slits. Notice that the index $l$ works as the label of a given 
slit displaced by $ld$ from the origin of the $\ce{D}$-slit aperture.
Inserting Eq.~(\ref{slits}) into Eq.~(\ref{Psi}) and considering 
a narrow pump beam transverse profile $W(\xi ;z_{A})$, which is 
 nonvanishing only inside a small interval $[-a,a]$, around the center 
($\xi =0$) of the plane $z=z_{A}$ (this can be achieved by focusing the pump 
beam), we can show that \cite{artleo}
\begin{equation}        \label{qudits}
\ket{\Psi} = \frac{1}{\sqrt{\ce{D}}} \sum_{l=-l_{D}}^{l_{D}}
          e^{ik\frac{d^{2}l^{2}}{2z_{A}}} \;
          \ket{l}\lsub{1} \otimes \ket{-l}\lsub{2},
\end{equation}
where
\begin{equation}      \label{base}
\ket{l}\lsub{j} \equiv \sqrt{\frac{a}{\pi}}
                \itgf{q_{j}} e^{-iq_{j}ld}\sinc(q_{j}a)\ket{1q_{j}}.
\end{equation}
The $\{\,\ket{l}\lsub{j}\}$ states  satisfy the condition
$\lsub{j}\braket{l}{l'}\lsub{j}=\delta_{ll'}$
and represent the photon in mode $j$ transmitted by the slit
$l$. We use them to define the logical states of the qudits.
Therefore, the two-photon state in Eq.~(\ref{qudits}) has the form of a 
maximally entangled state of two qudits having  correlations such that
if the photon in mode 1 passes through the slit $l$ the photon in mode 2
will pass through the symmetrically opposite slit $-l$.
These correlations satisfy the transverse momentum conservation in the thin
crystal approximation \cite{monken1}.
The relative phases are due to the optical path length
difference of the biphoton from the crystal to a given pair of slits.

\begin{figure}[tbh]
\begin{center}
\includegraphics[height=70mm,width=60mm]{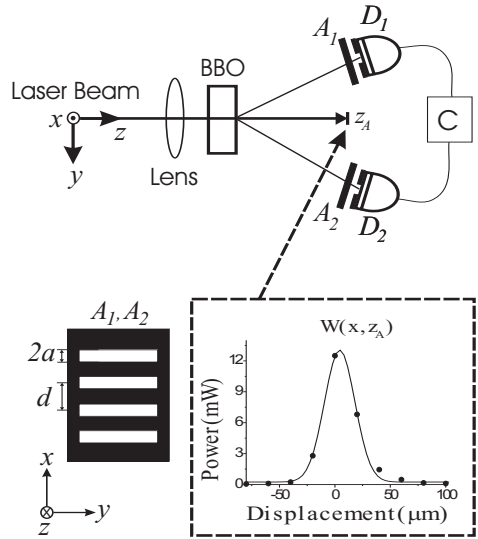}
\end{center}                
\caption{\label{fig:setup}  Outline of the experimental setup. The
inset shows the transverse profile of the pump beam in the $\ce{D}$-slits
plane, used in the measurements.}
\end{figure}

With the experimental setup outlined in Fig.~\ref{fig:setup} we make
measurements for $\ce{D}=4$ and 8.  A 5-mm-long BBO
($\beta$-barium borate) crystal, cut for type-II phase matching is pumped
by a 100~mW, frequency-doubled  Ti: sapphire laser operating at 413~nm.
Down-converted photons with a degenerate wavelength
$\lambda = 826$~nm are produced at an angle of
$2.5^{\circ}$ off the pump direction.
Two identical $\ce{D}$-slits ($A_{1}$ and $A_{2}$) are placed at the exit
path of the down-converted beams at the same distance $z_{A}= 200$~mm
from the crystal. The slit width is $2a=0.09$~mm and
the distance between two consecutive slits is $d=0.17$~mm.
Single and coincidence counts are measured
by detectors $D_{1}$ and $D_{2}$
placed behind the $\ce{D}$-slits at a distance $z\approx 2$~mm.
In front of each detector there is a single slit of width 0.1~mm
(oriented parallel to the slits of $A_{i}$),
followed by an interference filter of
8~nm bandwidth, centered at 826~nm.
A lens of $f=250$~mm is placed in the Gaussian pump beam at 50~mm before the
crystal such that it is focused in the plane of the
$\ce{D}$-slits, as shown in the inset of Fig.~\ref{fig:setup}.
This transverse profile ensures that the 
state in Eq.~(\ref{qudits}) will be formed after the  $\ce{D}$-slits.

In order to identify this state we make a coincidence selective measurement
onto the basis $\{\ket{l}\lsub{1}\ket{l'}\lsub{2}\}$ as follows: 
detector $D_{1}$ is kept fixed behind the slit $l$ while detector $D_{2}$ 
is scanning the other $\ce{D}$-slit in the $x$ direction. After $\ce{D}$ 
measurements like this 
($D_{1}$ ranging from $l=\frac{-(\ce{D}-1)}{2}$ to $l=\frac{\ce{D}-1}{2}$)
we will have the probability amplitudes for all $\ce{D}^{2}$ basis states
$\{\ket{l}\lsub{1}\ket{l'}\lsub{2}\}$. According to the quantum state in 
Eq.~(\ref{qudits}) each such measurement will select only 
$\ket{l}\lsub{1}\ket{-l}\lsub{2}$. Thus, it is expected that coincidences 
occur only when the detector $D_{2}$ scan the slit $l'=-l$. 
No diffraction or interference is observed in this process, because the 
detectors are very close to the slits (Fig.~\ref{fig:setup}).

\begin{figure}
\begin{center}
\includegraphics[width=0.45\textwidth]{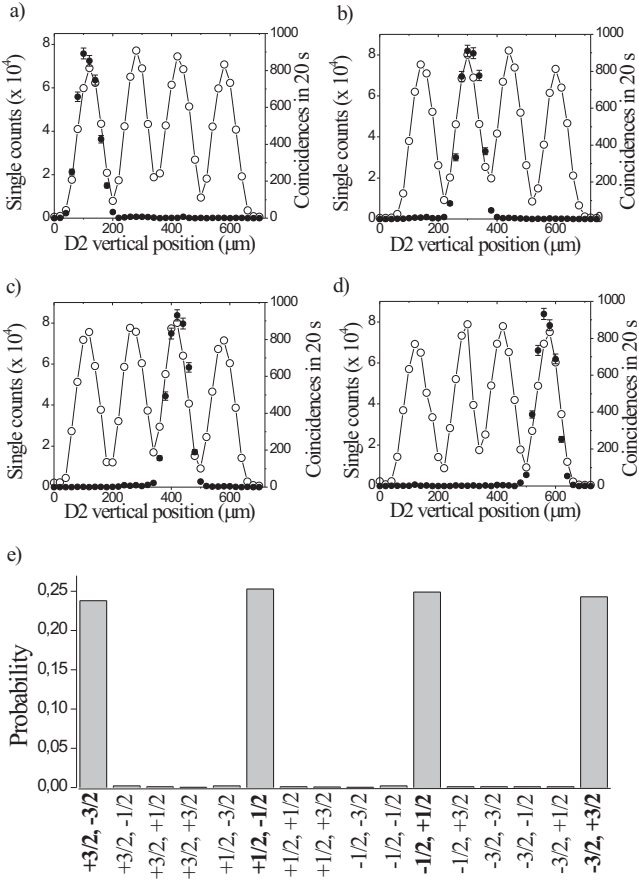}
\end{center}
\caption{\label{fig:ququart} $\ce{D}=4$. $D_{2}$ single counts
($\circ$) and $D_{1}$-$D_{2}$ coincidence counts  ($\bullet$)
measured simultaneously with $D_{1}$ fixed behind the slit $l$ and
$D_{2}$ scanning in $x$ direction. From left to right the single
count peaks are the slits $l'=-\frac{3}{2},\ldots,+\frac{3}{2}$.
 $D_{1}$ is fixed behind the slit $l$ (a) $\st +\frac{3}{2}$,
(b) $\st +\frac{1}{2}$, (c) $\st -\frac{1}{2}$, and (d) $\st
-\frac{3}{2}$. (e) Histogram of probabilities for all basis states
$\{\ket{l}\lsub{1}\ket{l'}\lsub{2}\}$.}
\end{figure}

\begin{figure}
\begin{center}
\includegraphics[width=0.45\textwidth]{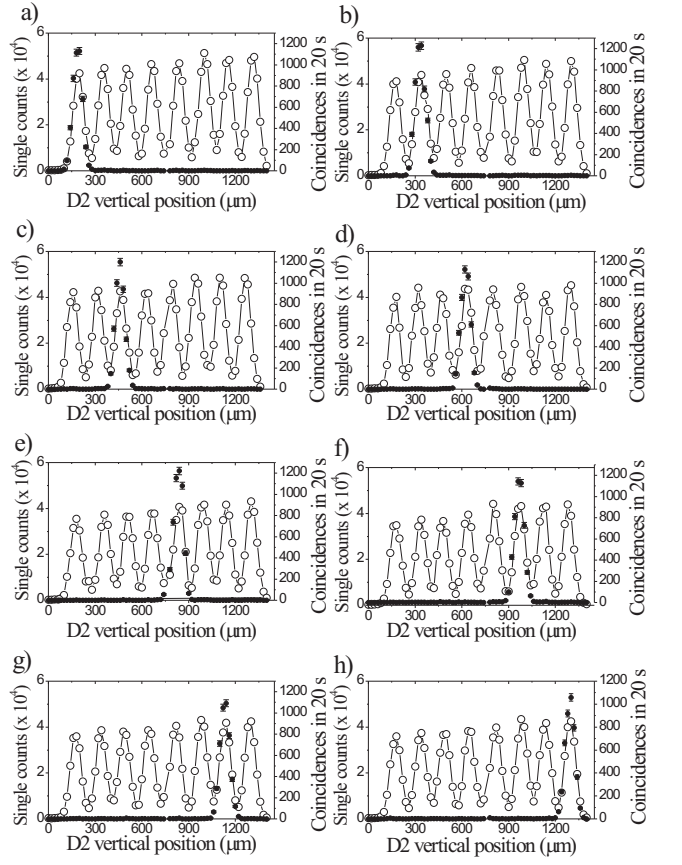}
\end{center}
\caption{\label{fig:f8} $\ce{D}=8$. $D_{2}$ single counts
($\circ$) and $D_{1}$-$D_{2}$ coincidence counts  ($\bullet$)
measured simultaneously with $D_{1}$ fixed behind the slit $l$ and
$D_{2}$ scanning in $x$ direction. From left to right the single
count peaks are the slits $l'=-\frac{7}{2},\ldots,+\frac{7}{2}$.
 $D_{1}$ is fixed behind the slit $l$ (a) $\st +\frac{7}{2}$,
(b) $\st +\frac{5}{2}$, (c) $\st +\frac{3}{2}$, (d) $\st +\frac{1}{2}$, 
(e) $\st -\frac{1}{2}$, (f) $\st -\frac{3}{2}$, (g) $\st -\frac{5}{2}$, 
and (h) $\st -\frac{7}{2}$.}
\end{figure}

The results for $\ce{D}=4$ and $\ce{D}=8$ are shown in
Figs.~\ref{fig:ququart} and \ref{fig:f8}, respectively. One can
see that the experimental  data are in agreement to what we
describe above: there are coincidence peaks only when the detector
$D_{2}$ scan the slit symmetrically opposite to that which
detector $D_{1}$ is fixed. It is also noteworthy that these peaks
have approximately the same number of coincidences within the
error bars. This means that the states $\ket{l}\lsub{1}\ket{-l}\lsub{2}$
will have the same amplitudes. To see this we calculate the
probabilities for all basis states. These ones are defined as
$P_{l_{1}l'_{2}}=C_{l_{1}l'_{2}}/\sum_{\{l_{1},l'_{2}\}}C_{\{l_{1}l'_{2}\}}$, 
where $C_{l_{1}l'_{2}}$ is the number of coincidences between detector
$D_{1}$ at slit $l$ and $D_{2}$ at $l'$ \cite{coment2}.
Figure~\ref{fig:ququart}(e)  shows a histogram of probabilities
for all basis states of $\ce{D}=4$. We clearly observe there, a
structure of 4-dimensional maximally entangled qudits. We observe
the same for $\ce{D}=8$. So within the experimental errors the
states $\ket{l}\lsub{1}\ket{-l}\lsub{2}$ are equally weighted. For $\ce{D}=4$
these errors are around 4\% and we obtain
\begin{eqnarray}
\ket{\Psi} & = &   0,50\;\ket{\st -\frac{1}{2}, +\frac{1}{2}}
      \; + \;  0,50\;\ket{\st +\frac{1}{2}, -\frac{1}{2}} \nonumber \\
  &  & \!\! \text{} +
     e^{i\frac{kd^{2}}{z_{A}}} (0,49\;\ket{\st -\frac{3}{2}, +\frac{3}{2}}
      \; + \;  0,49\;\ket{\st +\frac{3}{2}, -\frac{3}{2}}),
 \label{expquarts}
\end{eqnarray}
which has a fidelity $F=0.98\pm 0.08$ to that state predicted by theory in
Eq.~(\ref{qudits}). For $\ce{D}=8$, with errors around 3\%, we have
\begin{eqnarray}
\ket{\Psi} & = &   0,36\;\ket{\st -\frac{1}{2}, +\frac{1}{2}}
     \; + \;  0,34\;\ket{\st +\frac{1}{2}, -\frac{1}{2}} \nonumber \\
&  & \!\! \text{} +
  e^{i\frac{kd^{2}}{z_{A}}} (0,34\;\ket{\st -\frac{3}{2}, +\frac{3}{2}}
     \; + \;  0,34\;\ket{\st +\frac{3}{2}, -\frac{3}{2}})  \nonumber \\
&  & \!\! \text{} +
  e^{i\frac{3kd^{2}}{z_{A}}} (0,34\;\ket{\st -\frac{5}{2}, +\frac{5}{2}}
     \; + \;  0,36\;\ket{\st +\frac{5}{2}, -\frac{5}{2}}) \nonumber \\
&  & \!\! \text{} +
  e^{i\frac{6kd^{2}}{z_{A}}} (0,36\;\ket{\st -\frac{7}{2}, +\frac{7}{2}}
     \; + \;  0,35\;\ket{\st +\frac{7}{2}, -\frac{7}{2}}),
\label{expqu8}
\end{eqnarray}
with a fidelity $F=0.96\pm 0.05$ to that state in Eq.~(\ref{qudits}).
Phases in Eqs. (\ref{expquarts}) and (\ref{expqu8}) were not measured.
They are dependant on  fixed experimental parameters and they can be 
cancelled by choosing right values for $d$ and $z_{A}$ or by adding 
an appropriate external phase to a given slit.



Although the measurement procedure that we adopt demonstrates
that the correlations between the two photons agree with the theoretical
predictions, it is not sufficient
to guarantee that we have a coherent superposition of product states
$\ket{l}\lsub{1}\ket{-l}\lsub{2}$.
Exactly the same results shown in Figs.~\ref{fig:ququart} and \ref{fig:f8}
could be obtained by an equally weighted mixture like
\begin{equation}        \label{classico}
\rho_{\text{cc}} = \frac{1}{D} \sum_{l=-l_{D}}^{l_{D}}
       \ket{l}\lsub{1\,1\!}\!\bra{l} \otimes \ket{-l}\lsub{2\,2\!\!}\bra{-l}.
\end{equation}
This state, also called classically correlated, exhibits the same pairwise
correlation that one in Eq.~(\ref{qudits}). If in each
measurement the detectors are selecting one pair of \emph{independent} 
spatial modes, then we should have the state $\rho_{\text{cc}}$.
Let us show that is not the case here, i.e., the prepared two-photon
state after the $\ce{D}$-slit is indeed a pure and entangled state.
We define $\mathbb{E}_{j}\equiv\op{E}^{(+)}(x_{j},z)$ where
$\op{E}^{(+)}(x_{j},z)$ is the field operator for the mode $j$ and $x$ is the
detection position in the plane $z$. So, the photodetection coincidence
rate $C(x_{1},x_{2})$ is proportional to
$\text{Tr}(\rho \mathbb{E}^{\dag}_{1}\mathbb{E}^{\dag}_{2}%
\mathbb{E}_{2}\mathbb{E}_{1})$ \cite{Scully}. 
We consider detectors placed at $z\gg z_{A}$ after the apertures 
(Fraunhofer regime). Experimentally, this can be achieved by inserting a 
convergent lens with focal length $f$, after the apertures,  in each arm of 
the twin photons at a distance  $z_{L}$ from the crystal. In this way, the 
electric field operator is \cite{monken1,ivan}
$$
\mathbb{E}  =  \itg{q}\itgf{q'}
\op{a}(q') e^{i\bm{\left(}qx - q^{2}\frac{(z-z_{L})}{2k} 
- q'^{2}\frac{(z-z_{A})}{2k} + f\frac{(q-q')^{2}}{2k}\bm{\right)}}.
$$
The coincidence rate for the state in Eq.~(\ref{qudits}) will be 
\begin{eqnarray}
 C(x_{1},x_{2}) & \propto & 
\ds\sum_{l=-l_{D}}^{l_{D}} \!\! V_{ll}(x_{1},x_{2}) +
2 \!\! \sum_{l=-l_{D}}^{l_{D}-1}\sum_{m=l+1}^{l_{D}}
\!\! V_{lm}(x_{1},x_{2}) \nonumber \\[2mm]
& & \times   
\cos\bm{(}\beta (l-m)\bm{[}x_{2}-x_{1} - (l+m)\phi\bm{]}
\bm{)},
\label{qprob}
\end{eqnarray}
where $\beta =kfd/[f^{2}-(z-z_{L}-f)(z-z_{A}-f)]$ and
$\phi =d[f^{2}-(z-z_{L}-f)(z+z_{A}-f)]/2fz_{A}$;
$ V_{lm}(x_{1},x_{2})  \equiv  
\prod_{(r=1,2)(s=l,m)}\!\!
\sinc\!\!\bm{\left(}a\beta[x_{r}+(-1)^{r}s\eta d]/d\bm{\right)}, 
$
are the single slit diffraction terms where $\sinc (x)\equiv (\sin x)/x$ 
and $\eta=(z-z_{L}-f)/f$. For a classically correlated state
$C(x_{1},x_{2})$ will be given only by the first sum in 
Eq.~(\ref{qprob}). We see that for an entangled state, the coincidence rate 
exhibits an interference pattern with conditional fringes~\cite{GHZ}
(the fringes depend on the position of a given detector) while for the
mixture we will have only single slit diffraction and no interference. 
So, when we treat spatial correlations of two photons, the observation of
a fourth-order interference pattern with conditional fringes is  a sufficient
signature for entanglement.

\begin{figure}[tbh]
\begin{center}
\includegraphics[width=0.3\textwidth]{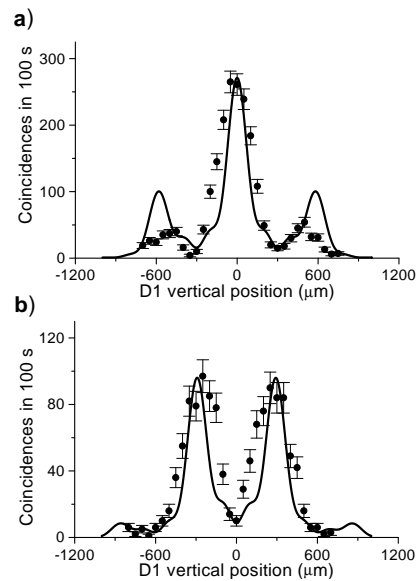}  
\end{center}
\caption{\label{interf}  Fourth-order interference patterns for two photons
transmitted by 4-slits as a function of $D_{1}$ position. 
(a) $D_{2}$ is kept fixed at $x_{2}=0\mu$m. 
(b) $D_{2}$ is kept fixed at $x_{2}=300\mu$m.}  \vskip -1.5mm
\end{figure}

Fourth-order interference patterns were obtained for $\ce{D}=4$. 
With the same setup shown in  Fig.~\ref{fig:setup}, we put the detectors 
at $z=800$mm from the crystal and convergent lenses with focal length 
$f=150$mm were inserted in each arm of the twin photons at $z_{L}=650$mm 
from the crystal.  Coincidence measurements were made as a function of
$x$ position of the detector $D_{1}$ while $D_{2}$ is kept fixed.
The results are shown in Fig.~\ref{interf}: (i)
$D_{2}$ fixed at $x_{2}=0\mu$m [Fig.~\ref{interf}(a)]; (ii)
$D_{2}$ fixed at $x_{2}=300\mu$m [Fig.~\ref{interf}(b)].
We clearly observe a two-photon interference pattern with conditional 
fringes. The solid curves fitting the experimental data, are obtained from 
the theoretical expression in Eq.~(\ref{qprob}) with only one normalization
parameter and taking into account the finite size of the detectors. 
The discrepancy between experimental data and theory  is likely due to 
the small acceptance angle of the detectors. 
Nevertheless, the main result shown in Fig.~\ref{interf} is the existence 
of a two-photon interference pattern which has conditional fringes. This
agree with the coincidence rate in Eq.~(\ref{qprob}) and
as we said, for spatially correlated photon pairs, this is a signature for 
entanglement.  Therefore, as we wish demonstrate the two-photon state 
prepared in our experiment is a pure and entangled state
 of qudits with $\ce{D}=4$ and $\ce{D}=8$.

We have presented a scheme to generate maximally entangled states of 
qudits by using two photons produced by SPDC and exploring the transverse
spatial correlation between them. The generated state is seen to
have a high fidelity to that predicted by theory and we show also that
it is indeed a pure and entangled state. This experiment could be used for 
tests on the foundations of quantum mechanics. In addition, there is also 
the possibility of carry out this experiment with  the $\ce{D}$-slits
replaced by optical fibers. This makes possible the construction of
 quantum-optical logic gates to achieve protocols in quantum communication
that need higher dimensional entangled states.

\begin{acknowledgments}
We acknowledge S. P. Walborn and L. Davidovich for very useful discussions.
This work was supported by the Brazilian agencies CAPES, CNPq, and 
Mil\^enio-Informa\c{c}\~ao Qu\^antica. 
Saavedra was supported by Grants Nos. FONDECYT 1010010 Milenio 
ICM P02-49F.
\end{acknowledgments}


\begin{thebibliography}{Burlakov(2002)}
\bibitem{coment2} Each $C_{l_{1}l'_{2}}$ is the number of coincidences in
four points of maximum single counts of the scanned slit $l'$.
\end{thebibliography}
\end{document}